\documentstyle[psfig]{aa}

\begin{document}

\topmargin 0mm
\title{Profiles of thermal line emission from advection dominated accretion
flows}

   \author{Xinwu Cao\inst{1}, Y.D. Xu\inst{2}, J.H. You\inst{2}}

\institute{
{\inst{1} Shanghai Astronomical Observatory, Chinese Academy of Sciences, 80 Nandan Road,
Shanghai, 200030, China, and Beijing Astrophysical Center(BAC), China}\\
{\inst{2}  Department of Applied Physics, Shanghai Jiaotong
University, Shanghai, 200030, China  } \\ }

\date{Received 26 January 2000/Accepted 11 July 2000}
\thesaurus{12(02.01.2; 02.12.3; 13.25.3)}

\offprints{Xinwu Cao (cxw@center.shao.ac.cn)}

\maketitle
\markboth
{Cao et al.: Profiles of thermal line emission from advection
dominated accretion flows}
{Cao et al.: Profiles of thermal line emission from advection
dominated accretion flows}

\begin{abstract}
  
Recently, Narayan \& Raymond (1999) proposed that the thermal emission
lines from the hot plasma in advection dominated accretion flows
(ADAFs) are potentially observable with the next generation of
X-ray observatories, with which the physical properties of some
X-ray sources can be probed. In ADAFs, the temperature of the ion is so
high that the thermal broadening of the line is important. We calculate
the profiles of thermal line emission from ADAFs, in which both the
thermal and Doppler broadening have been considered. It is found that
the double-peaked profiles
are present for high inclination angles between the axis of disk and the
line of sight. The double-peaked profiles are smeared in low inclination
cases, and completely disappear while the inclination angle is less
than $15^{\circ}$,  where the thermal and turbulent broadening
dominated on the line profiles. We also note that the thermal 
line profile is affected by the location of the transition radius 
of ADAF.  The self-similar height-integrated disk structure 
and the emissivity with power-law dependence of radius are adopted 
in our calculations. The results obtained in this work can be used as a
diagnosis on the future X-ray observations of the thermal lines.
Some important physical quantities of ADAFs could 
be inferred from future thermal line observations.

\end{abstract}


\section{Introduction}

Emission line profiles from accretion disks have been investigated
by many researchers from both observational and theoretical sides.
In the past decade, the X-ray line emission has been extensively
studied, mainly stimulated by the progresses on X-ray
observations. The
profiles of emission line from accretion disks around a Schwarzschild
black hole (Fabian et al., 1989; Matt et al., 1993), or around a Kerr
black hole (Laor 1991) are well calculated.
The line spectra from ADAFs have been calculated by
Fukue \& Ohna (1997) without considering the thermal broadening of line
profile in the rest frame of the emitter.
The fluorescent line emission
from cool gas irradiated
by hot corona has been studied in some detail (Mushotzky et al., 1993;
Tanaka et al.,1995), which is the focus in most X-ray line formation 
mechanism investigations. The fluorescent line can be formed in the disk region very close
to the black hole. However, the thermal X-ray line emission from hot,
optically thin gas around black holes has not been discussed very much.
More recently, Narayan \& Raymond (1999) present the calculations on thermal
line formation in hot ADAFs, and
propose that such X-ray emission lines are potentially observable with
the next generation of X-ray observatories. The physical properties
of some X-ray sources and the constraints on the accretion flows in these
sources may be provided by future observatories.

In a two-temperature ADAF, while the ions receive most of the viscous energy
and are nearly virial, the electrons are heated
by several processes, such as, Coulomb coupling with the ions, compression,
and direct viscous heating, and cooled by a variety of radiation processes
(Narayan \& Yi, 1995b). The temperature of the electron is in general lower
than that of the ions due to insufficient interaction between electrons and
ions(see Narayan et al. 1998 for a review on ADAFs). The thermal X-ray line emission
is supposed to be from the disk region where the electron temperature less
than $10^{9}$ K, which corresponds to the disk region $r>100r_{g}$(Narayan \&
Raymond 1999). The profiles of the fluorescent line are mainly determined by
the Doppler broadening caused by the bulk motion of the flows, and the
thermal broadening can be neglected in the
profile calculations. Unlike the case of the fluorescent line, the thermal
X-ray lines are supposed to be from the disk region $r>100r_{g}$ with the
ion temperature
around $10^{10}$ K. In this case, the thermal broadening of the line profiles
is important compared with the Doppler broadening, since the bulk 
velocity of the ADAFs is only a fraction of the Kepler velocity, and
the Kepler velocity at $100r_{g}$ is only one-fifth of that at the inner
edge of the disk around a Schwarzschild black hole. In this work, we
present calculations on the thermal line profiles from ADAFs. In Sects.
2-4, we give the formalism of the problem. The last section contains
the results and a discussion.

\section{Thermal broadening of the line profiles}

We first calculate the thermal broadening of line emission from relativistic
ions. Using the common relativistic transformations and assuming the 
motion of thermal ions to be isotropic, we write the observed flux
density of the line from the ions moving at $\beta_{i}=v_{i}/c$ as
\begin{equation}
I_{\nu}(\gamma_{i}, ~\nu)=
{\frac {\epsilon_{\rm line}}
{2\beta_{i}\gamma_{i}^{3}(1-\beta_{i}\mu)^{2}\nu_{e}}},
\end{equation}
where $\epsilon_{\rm line}$ is the line emission power in the rest frame of
the ion, $\nu$, $\nu_{e}$ are the frequency measured by the observer 
and the line frequency measured in the emitter's frame respectively, $\gamma_{i}$ is the Lorentz
factor of the ion's thermal motion, and $\mu=\cos\theta$ ($\theta$ is the
angle between the motion of the ion and the line of sight).
Using the relation $\nu_{e}=\nu\gamma_{i}(1-\beta_{i}\mu)$, we can rewrite
Eq. (1) as
\begin{equation}
I_{\nu}(\gamma_{i}, ~\nu)=
{\frac {\epsilon_{\rm line}\nu^2} {2\beta_{i}
\gamma_{i}\nu_{e}^3}},
\end{equation}
where
\begin{displaymath}
{\frac {\nu_e}{\gamma_{i}(1+\beta_{i})}}\le \nu \le
{\frac {\nu_e}{\gamma_{i}(1-\beta_{i})}}.
\end{displaymath}
If we consider isotropic Maxwelliam distribution of the ions, the number
density per unit energy is
\begin{equation}
N(\gamma_{i})=\frac {1}{\Theta_{\rm ion}K_{2}({1/\Theta_{\rm ion}})}\gamma_{i}(\gamma_{i}
^{2}-1)^{1\over 2}e^{-\gamma_{i}/ \Theta_{\rm ion}},
\end{equation}
where $\Theta_{\rm ion}$ is the thermal temperature in units of ion rest
mass, and $K_{2}$ is the second-order modified Bessel function. The 
ion distribution function is normalized as 
\begin{displaymath}
\int\limits_{1}^{\infty} N(\gamma_{i}) d\gamma_{i}=1.
\end{displaymath}
We can therefore obtain the broadening profile of the thermal line emission as
\begin{displaymath}
I_{\nu}(\Theta_{\rm ion}, ~\nu)=\int N(\gamma_{i})d\gamma_{i}
{\frac {\epsilon_{\rm line}\nu^2} {2\beta_{i}
\gamma_{i}\nu_{e}^3}}
\end{displaymath}
\begin{equation}
={\frac {\epsilon_{\rm line}\nu^{2}}{2\Theta_{\rm
ion}K_{2}({1/\Theta_{\rm ion}})
\nu_{e}^{3}}}\int\limits_{\gamma_{\rm min}}^{\infty}\gamma_{i}
e^{-\gamma_{i}/ \Theta_{\rm ion}}d\gamma_{i},
\end{equation}
where 
\begin{displaymath}
\gamma_{\rm min}=
{\frac {\nu}{2\nu_e}}
\left[\left({\frac {\nu_e}{\nu}}\right)^{2}+1\right].
\end{displaymath} 

For the given ion temperature $\Theta_{\rm ion}$, we can integrate Eq. (4)
and obtain the line
profile:
\begin{displaymath}
I_{\nu}(\Theta_{\rm ion}, ~\nu)=
{\frac {\epsilon_{\rm line}\nu^{2}}{2K_{2}({1/\Theta_{\rm ion}})
\nu_{e}^{3}}}
\end{displaymath}
\begin{equation}
\times\left\{{\frac {\nu}{2\nu_e}}
\left[\left({\frac {\nu_e}{\nu}}\right)^{2}+1\right]
+\Theta_{\rm ion}\right\}\exp\left\{
{-{\frac {\nu}{2\nu_{e}\Theta_{\rm ion}}}
\left[\left({\frac {\nu_e}{\nu}}\right)^{2}+1\right]}\right\}. 
\end{equation}

Besides the thermal broadening, the turbulent velocity of eddies 
in ADAFs may also affect the line profile. 
We assume the turbulence to be isotropic and adopt a mean 
turbulent velocity $v_t=\beta_t c$. The
thermal line profile can be calculated 
by
\begin{displaymath}
I_{\nu}(\gamma_t, ~\Theta_{\rm ion}, ~\nu)=
{\frac {\epsilon_{\rm line}\nu^2}
{4\beta_t\gamma_t K_2(1/\Theta_{\rm ion})\nu_e^3}}
\end{displaymath}
\begin{equation}
\times\int\limits_{x_{\rm min}}^{x_{\rm max}} 
\left ({{1+x^2}\over {2x^2}}+{{\Theta_{\rm ion}}\over x}\right)
\exp\left(-{{1+x^2}\over {2\Theta_{\rm ion} x}}\right)dx,
\end{equation}
where 
\begin{displaymath}
x_{\rm min}=\gamma_t(1-\beta_t){\nu\over {\nu_e}},
\end{displaymath}
and 
\begin{displaymath}
x_{\rm max}=\gamma_t(1+\beta_t){\nu\over {\nu_e}}.
\end{displaymath}

\section{Structure of self-similar ADAF}

We consider the self-similar disk structure given by Narayan \& Yi (1994).
In this solution, the azimuthal rotational velocity $v_{\phi}$ and radial
velocity $v_{r}$ are given as a function of $r$ as follows:
\begin{equation}
v_{r}(r)=-\frac {5+3\epsilon^{\prime}}{3\alpha^{2}}g(\alpha,\epsilon^{\prime})
\alpha v_{K}(r)
\end{equation}
\begin{equation}
v_{\phi}(r)=\left[\frac {2\epsilon^{\prime}(5+2\epsilon^{\prime})}
{9\alpha^{2}}g(\alpha,
\epsilon^{\prime})\right]^{1/2}v_{K}(r)
\end{equation}
where,
\begin{displaymath}
v_{K}(r)=\left(\frac {GM}{r}\right)^{1/2},
\end{displaymath}
\begin{displaymath}
\epsilon^{\prime}\equiv {\frac {\epsilon}{f}}=\frac {1}{f}\left(\frac {5/3-\gamma}
{\gamma-1}\right),
\end{displaymath}
\begin{displaymath}
g(\alpha,~\epsilon^{\prime})\equiv \left[1+{\frac {18\alpha^{2}}
{(5+2\epsilon^{\prime})^{2}}}\right ]^{1/2}-1.
\end{displaymath}
It is assumed that magnetic fields contribute a constant fraction
$(1-\beta)$ of the total pressure in ADAFs. The appropriate relation
between the ratio of the specific heats $\gamma$ and $\beta$ is given
by Esin (1997): $\gamma=(8-3\beta)/(6-3\beta)$. 
The disk structure is then described by three parameters: the ratio of the
magnetic pressure $\beta$, the viscosity $\alpha$, and the fraction $f$ of viscously
dissipated energy which is advected.

For simplicity, we further assume that the emissivity $j$ of the unit area of
disk surface has power-law dependence of radius $r$,
\begin{equation}
j(r)\propto {r^{-b}}.
\end{equation}

The radial velocity $v_{r}$ and azimuthal velocity $v_{\phi}$ of the
flow given by Eqs. (7) and (8) will be used in the calculations of line
profile.

\section{Profiles of thermal lines from ADAFs}

Assuming the line of sight to be inclined at an angle $i$ to the disk axis,
the angle $\delta$ between the line of sight and the motion of gas in
the disk is
\begin{equation}
\cos\delta=\sin i \cos\left(\phi+\theta_{v}+\frac {\pi}{2}\right),
\end{equation}
where $\tan\theta_{v}=v_{r}/v_{\phi}$. 
Using the relativistic transformations, the line profile for a ring
in the
disk between  $r$ and $r+dr$ is available,
\begin{equation}
f_{\nu}(r, ~\nu_{obs})=rjdr \cos{i}\int {\frac {I_{\nu}(\Theta_{\rm ion},~
\nu^{\prime})}{\gamma^{3}_{b}(1-\beta_{b}\cos\delta)^{3}}} d\phi,
\end{equation}
where $\beta_{b}=\sqrt{v_{r}^{2}+v_{\phi}^{2}}/c$,
$\gamma_{b}=\sqrt{1-\beta_{b}^{2}}$, and $\Theta_{\rm ion}$ is
given by 
\begin{equation}
T_{i}=6.66\times 10^{12} {\frac {2(5+2\epsilon^{\prime})}
{9\alpha^2}} g(\alpha, \epsilon^{\prime})\beta
r^{-1}.
\end{equation}
The electron temperature $T_e$ has been neglected in Eq. (12) 
(see Eq. (2.16) in 
Narayan \& Yi 1995b), since $T_e$ is about an order of magnitude 
lower than $T_i$ in two-temperature ADAFs. 
In this paper, we focus our calculations mainly on iron, i.e., 
$\Theta_{\rm ion}={\frac {kT_i} {m_{Fe}c^2}}$. The relation
between $\nu^{\prime}$ and $\nu_{obs}$ is
\begin{equation}
\nu^{\prime}=\nu_{obs}\left(1-\frac {1}{r}\right)^{-1/2}\gamma_{b}(1-\beta_{b}\cos\delta)
\end{equation}
where the factor $(1-1/r)^{-1/2}$ represents  the gravitational
redshift. From Eqs. (10) and (13), we can obtain
\begin{displaymath}
\frac {d\phi}{d\nu^{\prime}}=
\end{displaymath}
\begin{equation}
\frac { \left(1-{\frac {1}{r}}\right)^{1/2} }{\nu_{obs}
\gamma_{b}
\beta_{b} \sin i \left \{1-\frac {1}{\beta^{2}_{b}\sin^{2}i}\left [1-\frac
{\nu^{\prime}}
{\gamma_b\nu_{obs}}(1-\frac {1}{r})^{1/2}\right ]^{2} \right \}^{1/2}},
\end{equation}

Finally, we can calculate the line profile as
\begin{displaymath}
F_{\nu}(\nu_{obs})=\int\limits_{r_{in}}^{r_{out}} rjdr \cos i
\end{displaymath}
\begin{equation}
\times\int
\limits_{\nu^{\prime}_{\rm min}}^{\nu^{\prime}_{\rm max}}
\frac {(\nu_{obs}/\nu^{\prime})^{3}\left(1-\frac {1}{r}\right)^
{-1}I_{\nu}
(\gamma_t, ~\Theta_{\rm ion}, ~ \nu^{\prime})d\nu^{\prime}}{\left 
\{\gamma_{b}^{2}\beta_{b}^{2}\nu_{obs}^{2}
\sin ^{2}i-\left [ \gamma_{b}\nu_{obs}
-\nu^{\prime}(1-\frac {1}{r})^{1/2} \right ]^{2}\right \}^{1/2} },
\end{equation}
where 
\begin{displaymath}
\nu^{\prime}_{\rm min}=\gamma_b\nu_{obs}
(1-\beta_b\sin i)
\left(1-{1\over r}\right)^{-1/2}
\end{displaymath}
and 
\begin{displaymath}
\nu^{\prime}_{\rm min}=\gamma_b\nu_{obs}
(1+\beta_b\sin i)
\left(1-{1\over r}\right)^{-1/2}.
\end{displaymath} 
The inner and outer radii of the line emission region are assumed
to be $r_{in}=100 r_g$ and $r_{out}=1000 r_g$. The typical turbulent 
velocity of eddies in ADAFs is approximated as $v_t\simeq \alpha c_s$, 
$c_s=\left({{kT_i}\over {\mu_i m_u}}\right)^{1/2}$, where  
$\mu_i=1.23$ is adopted, same as Narayan \& Yi (1995b).  Now, using 
Eqs. (6) and (15),
we can calculate the profiles of the thermal line from ADAFs.

\begin{figure}
\centerline{\psfig{figure=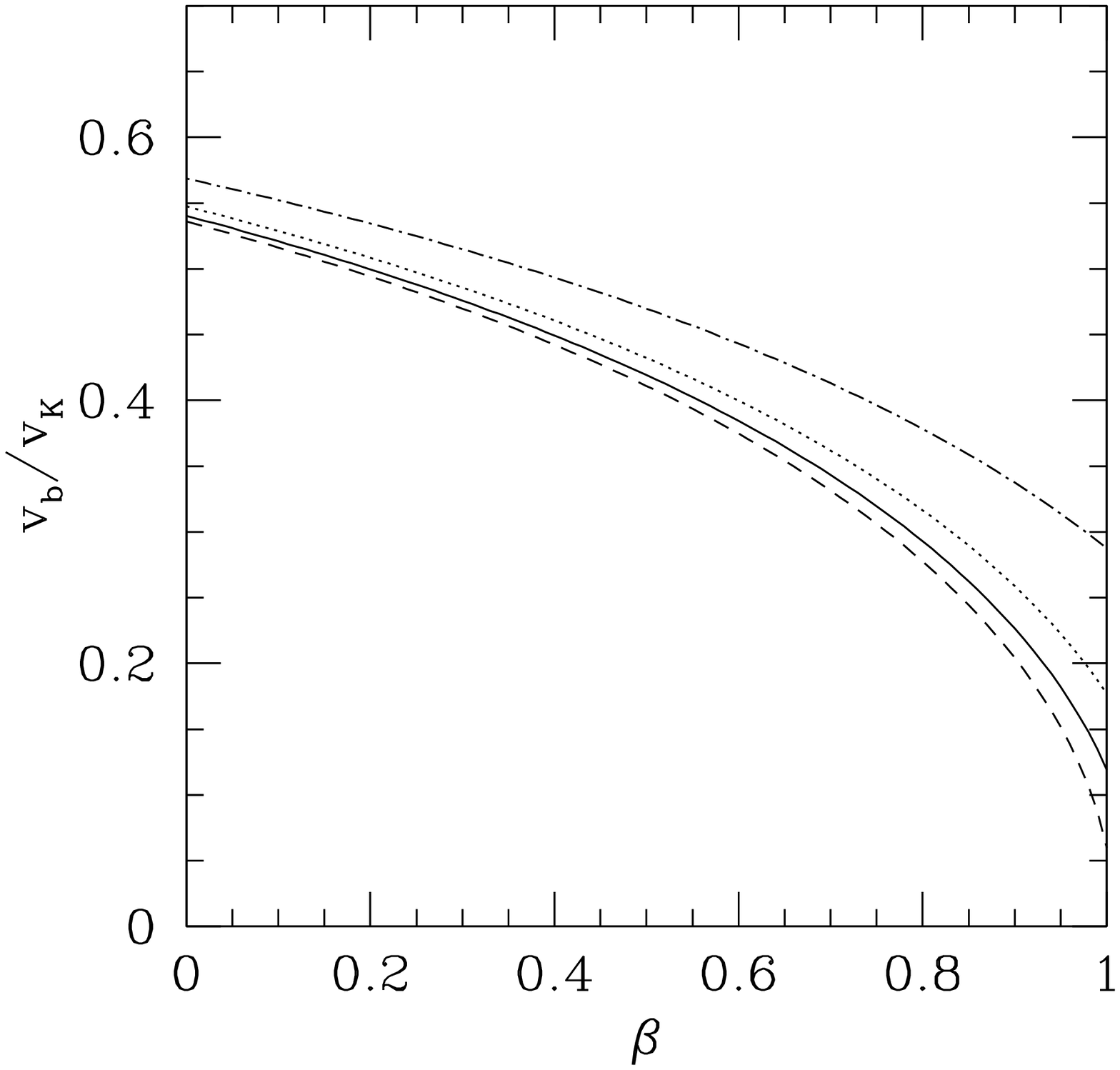,width=7.0cm,height=7.0cm}}
\caption{The velocity of the accretion flow as functions of $\beta$ for
different values of viscosity: $\alpha=$0.1 (dashed), 0.2(solid),
0.3(dotted), 0.5(dash-doted), respectively. }
\end{figure}

\begin{figure}
\centerline{\psfig{figure=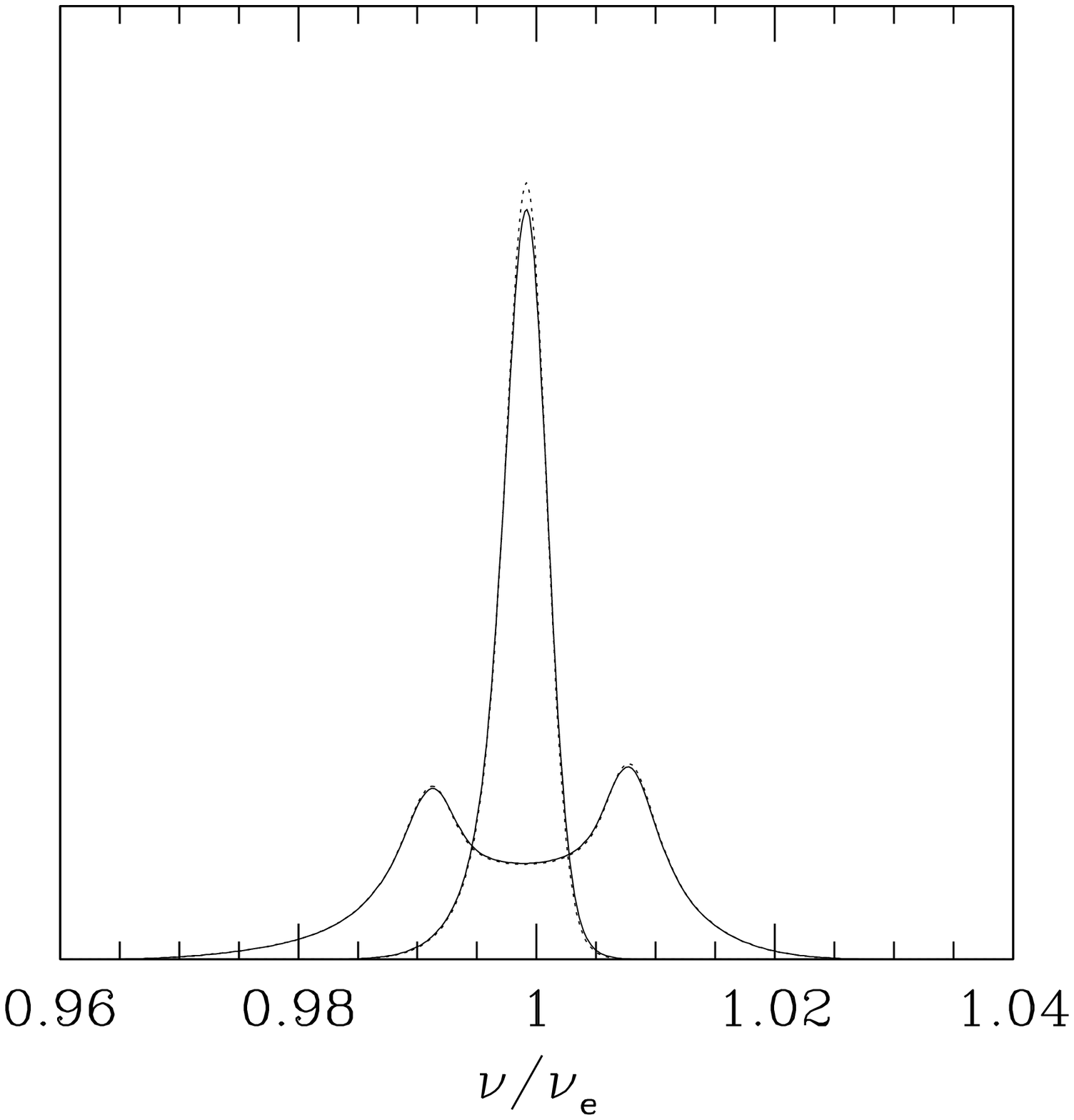,width=7.0cm,height=7.0cm}}
\caption{The thermal iron line profiles with parameters:  $\alpha=0.1$,
$\gamma=1.444$ ($\beta=0.5$), and 
$i=0^\circ$, $60^\circ$, respectively. The dotted lines indicate 
the profiles without considering turbulent motions in ADAFs. }
\end{figure}

\begin{figure}
\centerline{\psfig{figure=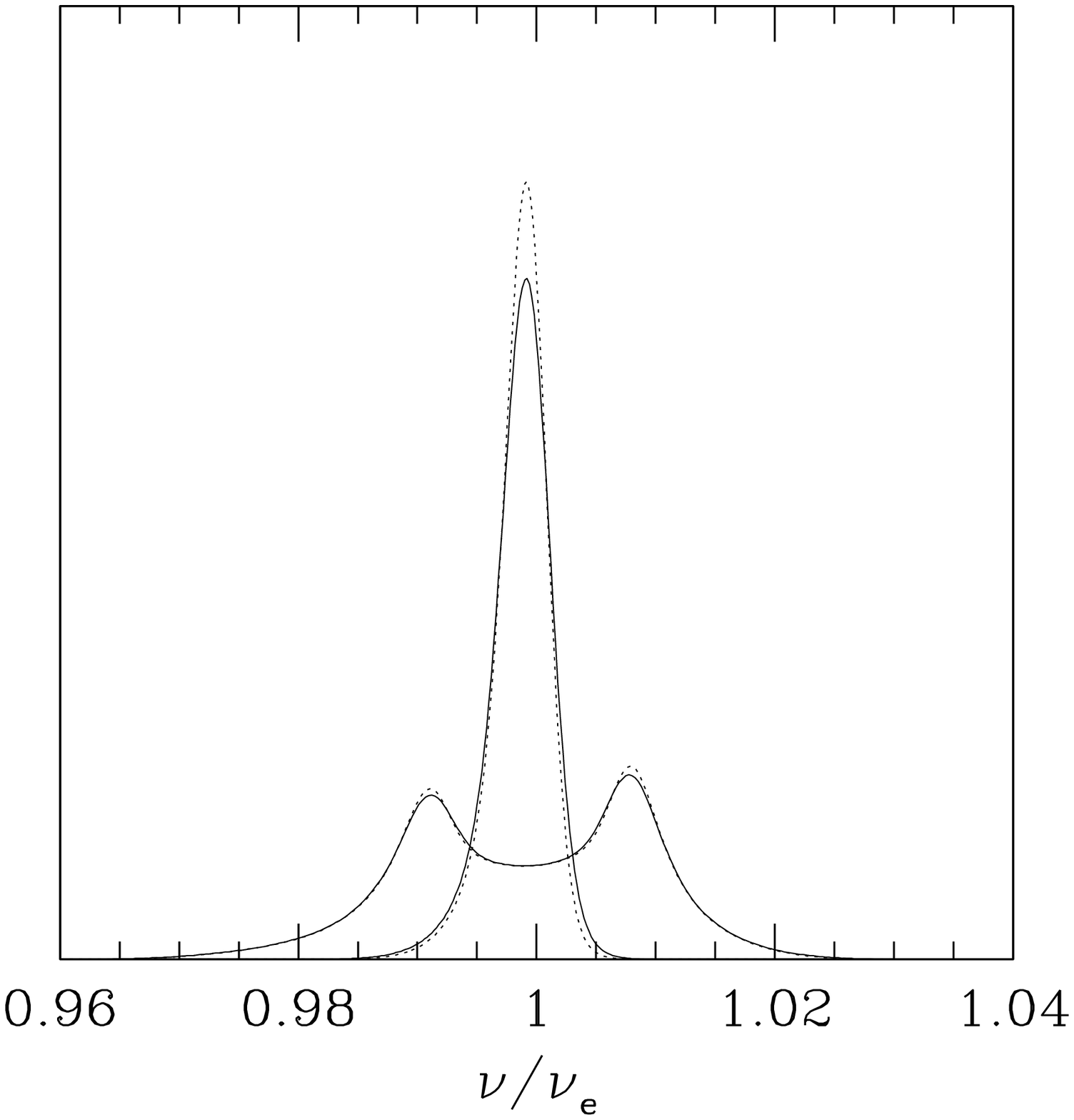,width=7.0cm,height=7.0cm}}
\caption{Same as Fig. 2, but $\alpha=0.2$. }
\end{figure}

\begin{figure}
\centerline{\psfig{figure=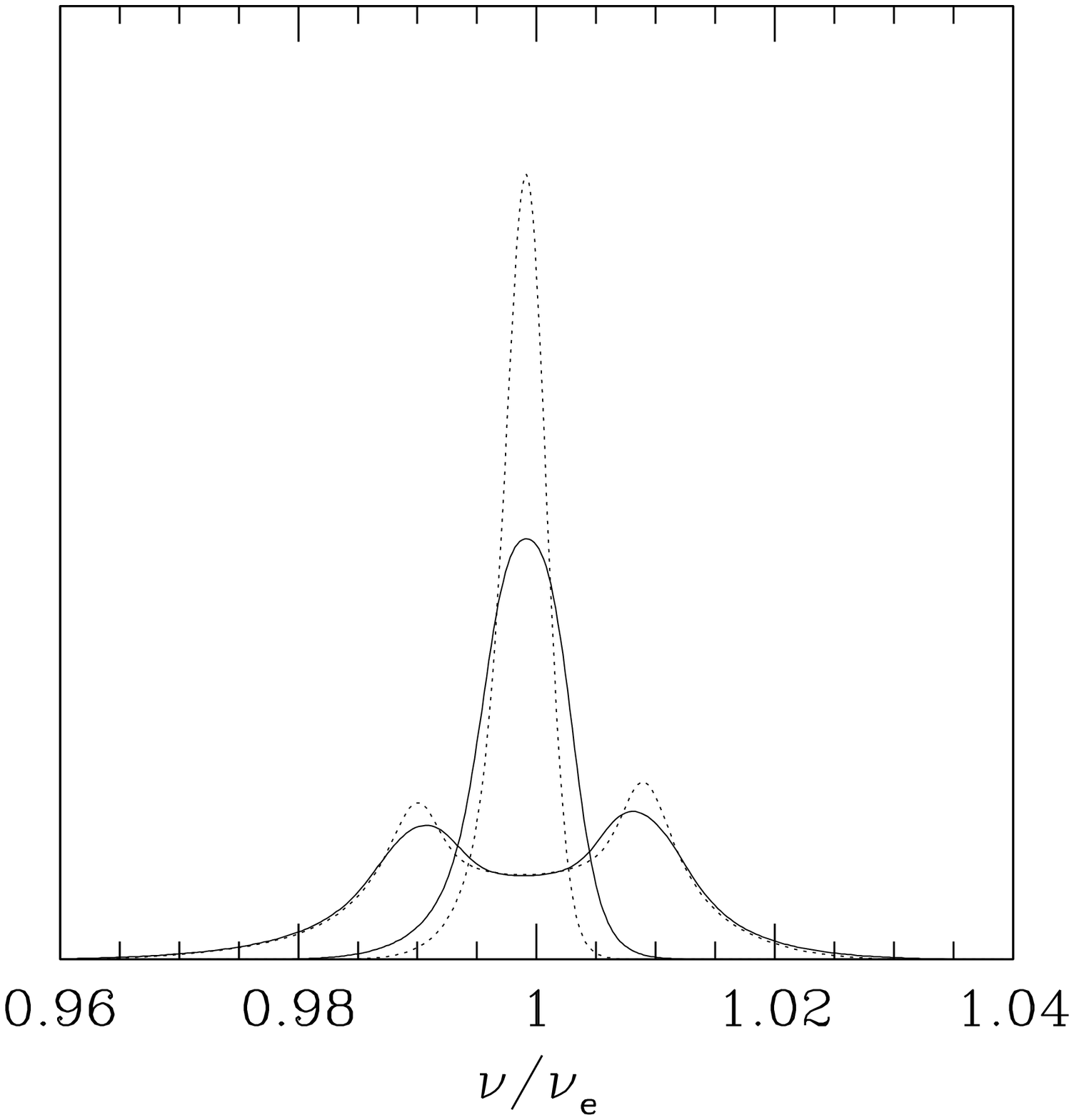,width=7.0cm,height=7.0cm}}
\caption{Same as Fig. 2, but $\alpha=0.5$. }
\end{figure}

\begin{figure}
\centerline{\psfig{figure=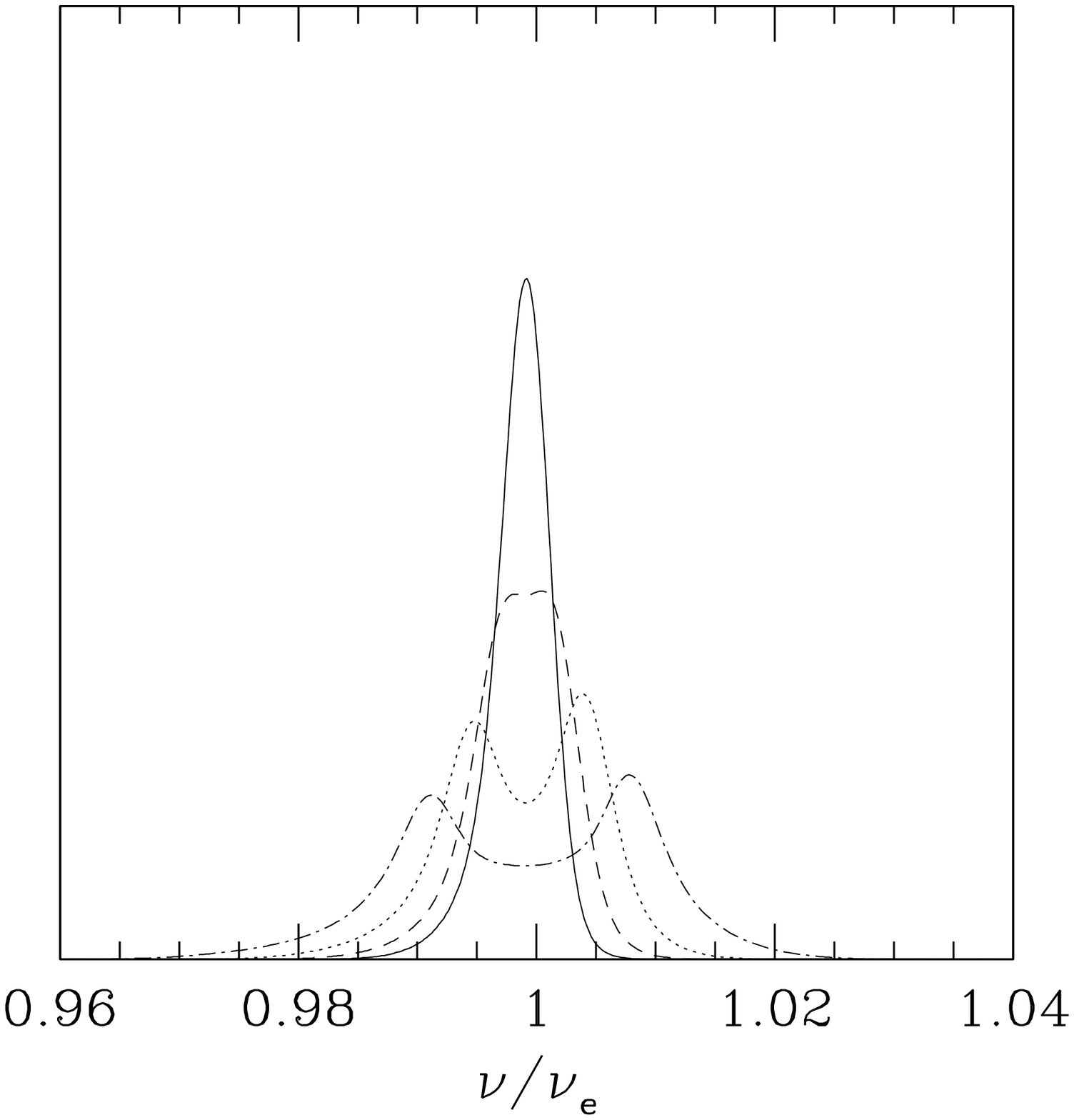,width=7.0cm,height=7.0cm}}
\caption{The profiles of the thermal iron line emission from accretion
flows for model A: $\alpha=0.2$, $f=1$, $\gamma=1.444$ ($\beta=0.5$), 
with different inclination
angles: $i=~0^\circ$(solid), 15$^\circ$(dashed), 30$^\circ$(dotted),
60$^\circ$(dash-dotted).}
\end{figure}

\begin{figure}
\centerline{\psfig{figure=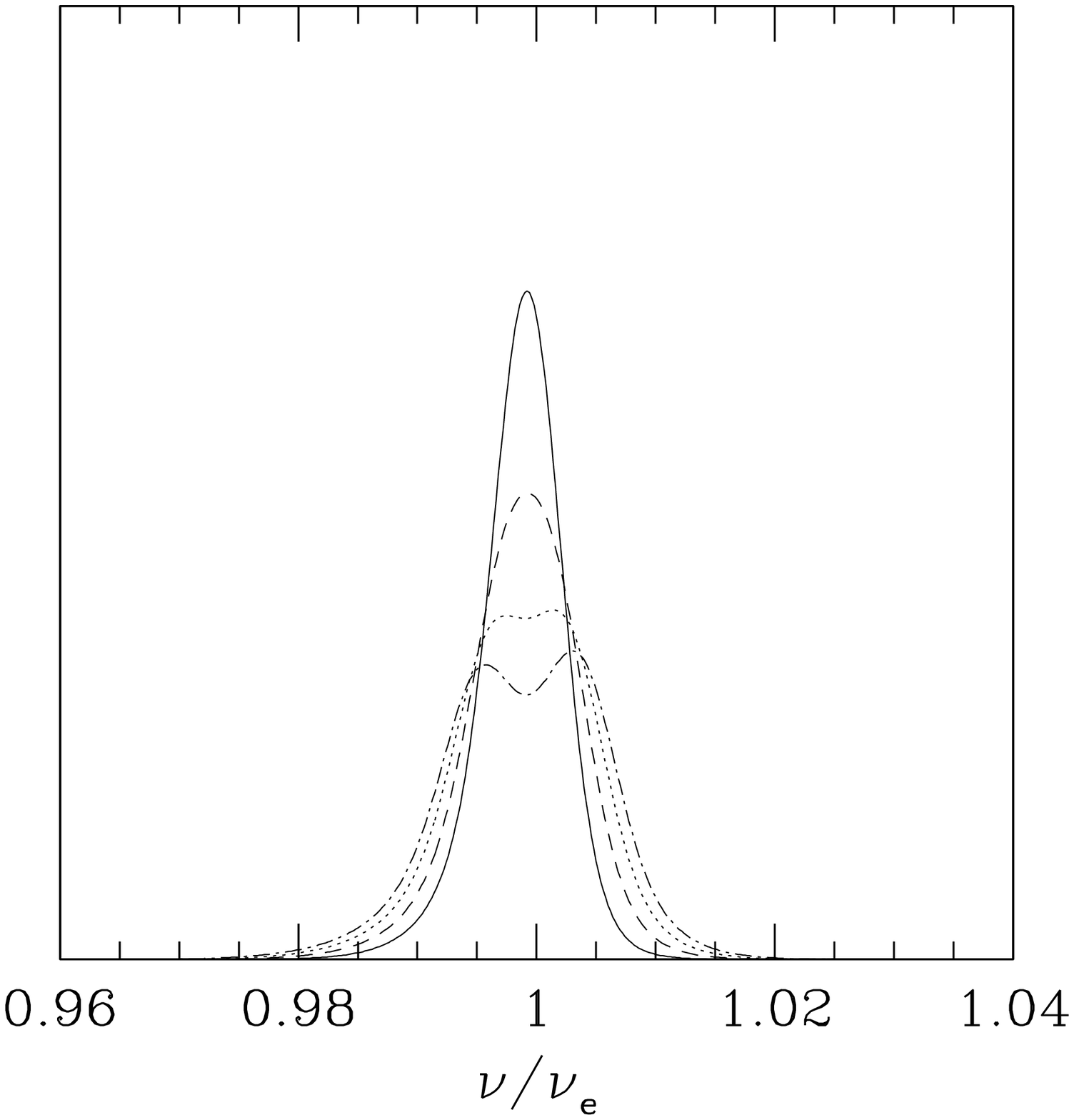,width=7.0cm,height=7.0cm}}
\caption{The profiles of the thermal iron line emission from accretion
flows for
model B: $\alpha=0.2$, $f=1$, $\gamma=1.602$ ($\beta=0.9$), with
different inclination
angles: $i=~0^\circ$(solid), 30$^\circ$(dashed), 45$^\circ$(dotted),
60$^\circ$(dash-dotted).}
\end{figure}

\begin{figure}
\centerline{\psfig{figure=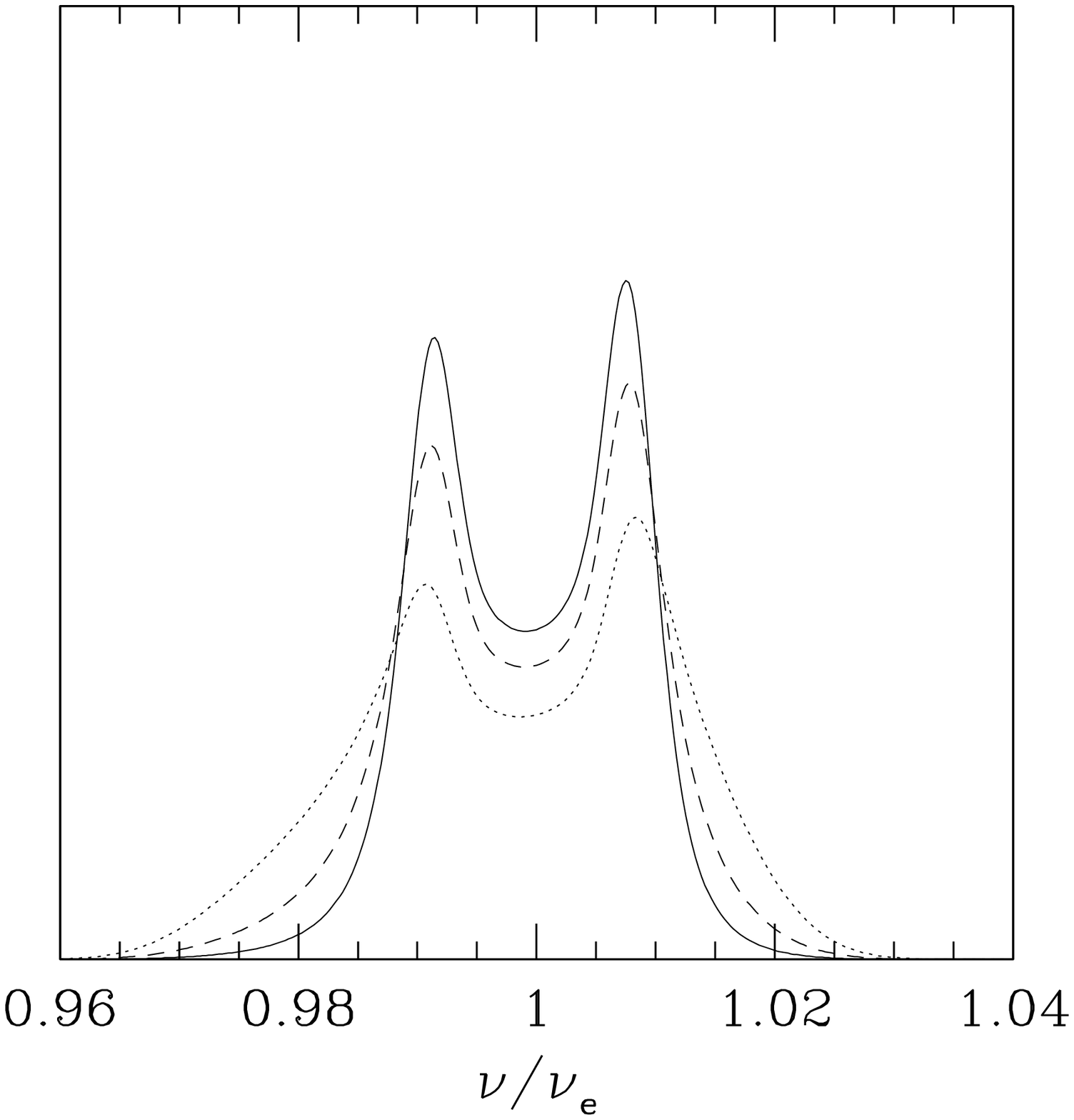,width=7.0cm,height=7.0cm}}
\caption{The profiles of thermal iron line emission from accretion flows for
model A with $i=60^\circ$ for different emissivity law: $b=~0$(solid),
1(dashed), 2(dotted).}
\end{figure}

\begin{figure}
\centerline{\psfig{figure=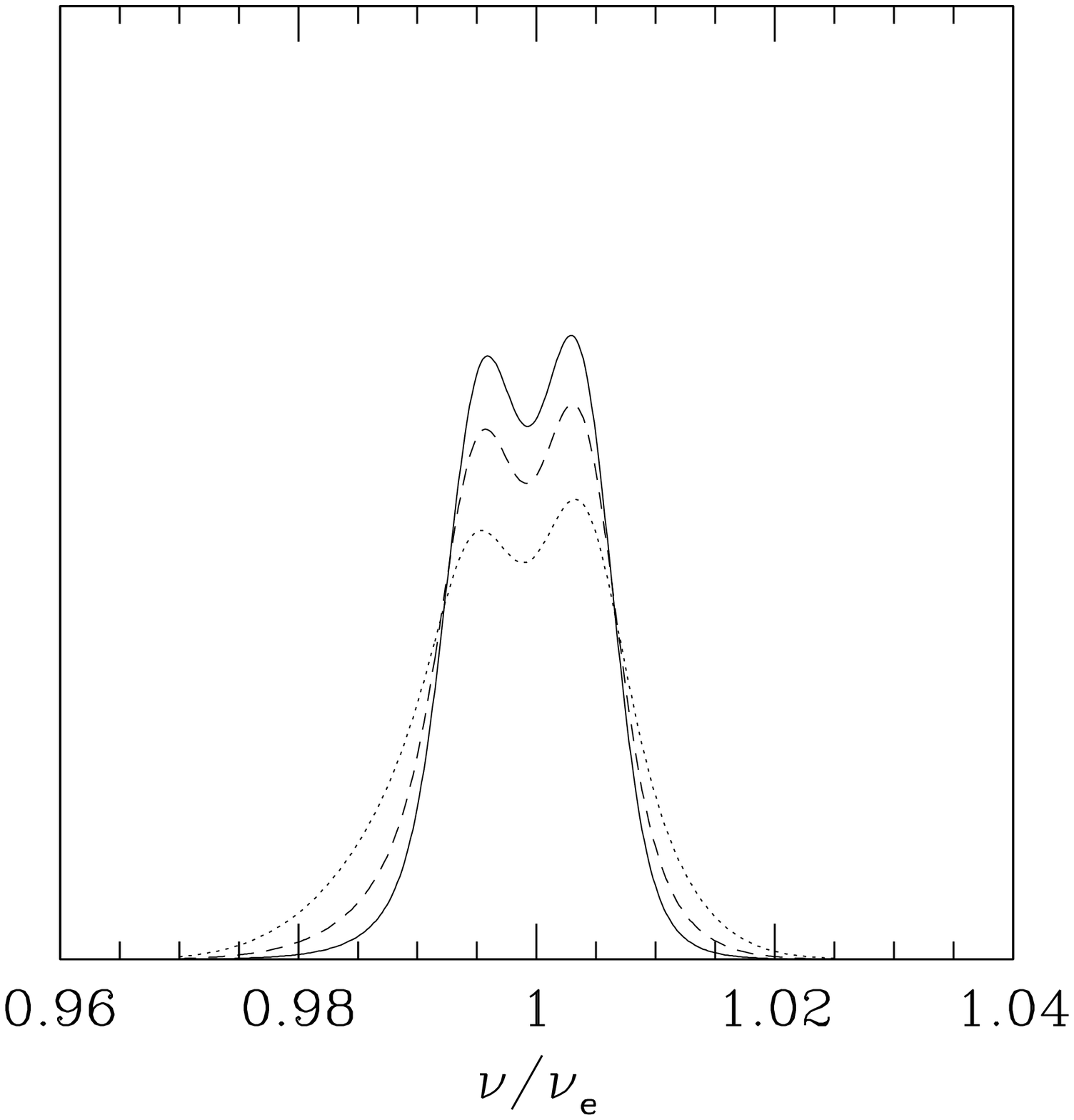,width=7.0cm,height=7.0cm}}
\caption{Same as Fig. 7, but for model B. }
\end{figure}

\begin{figure}
\centerline{\psfig{figure=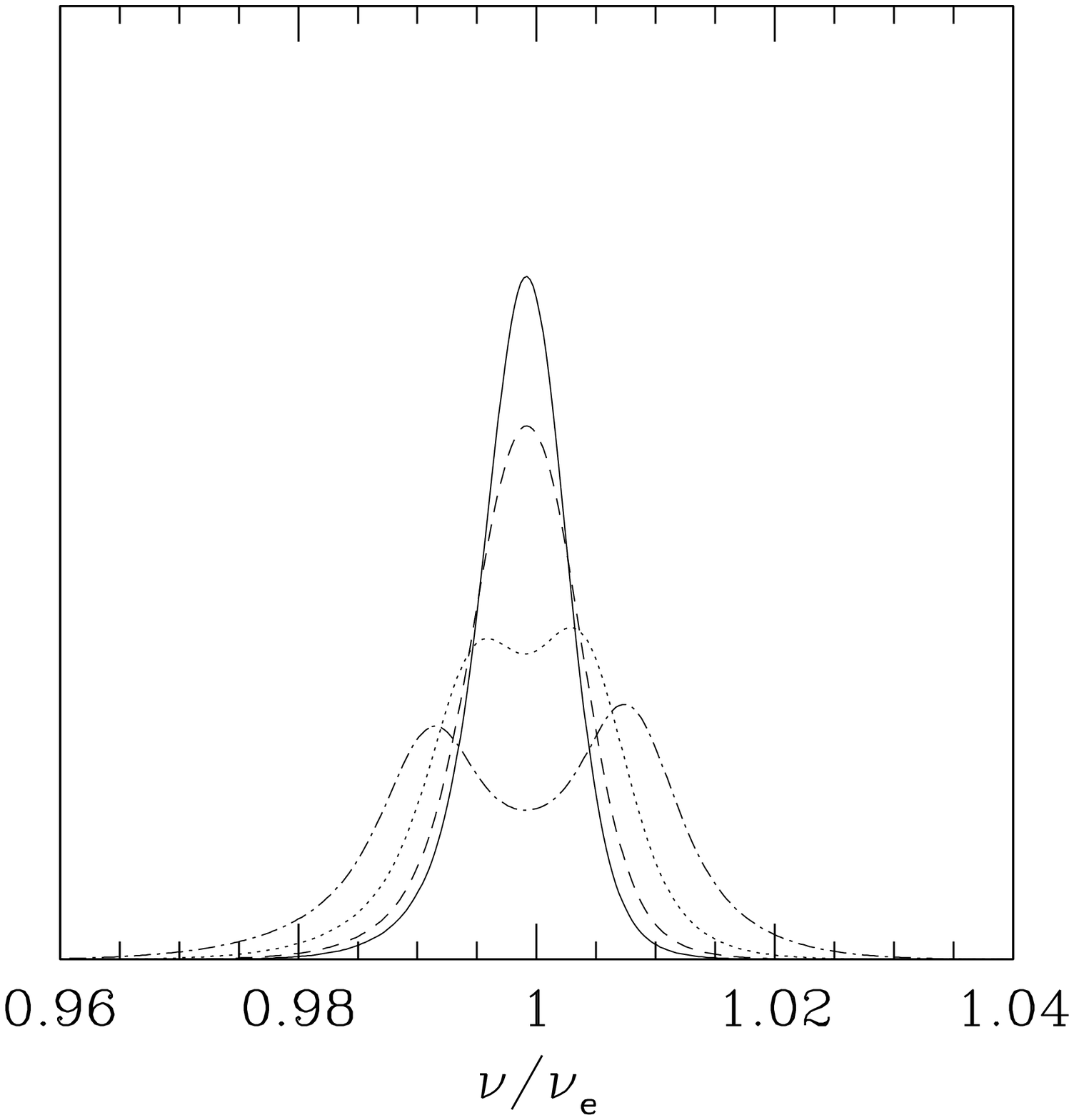,width=7.0cm,height=7.0cm}}
\caption{Same as Fig. 5, but for thermal oxygen lines. }
\end{figure}

\begin{figure}
\centerline{\psfig{figure=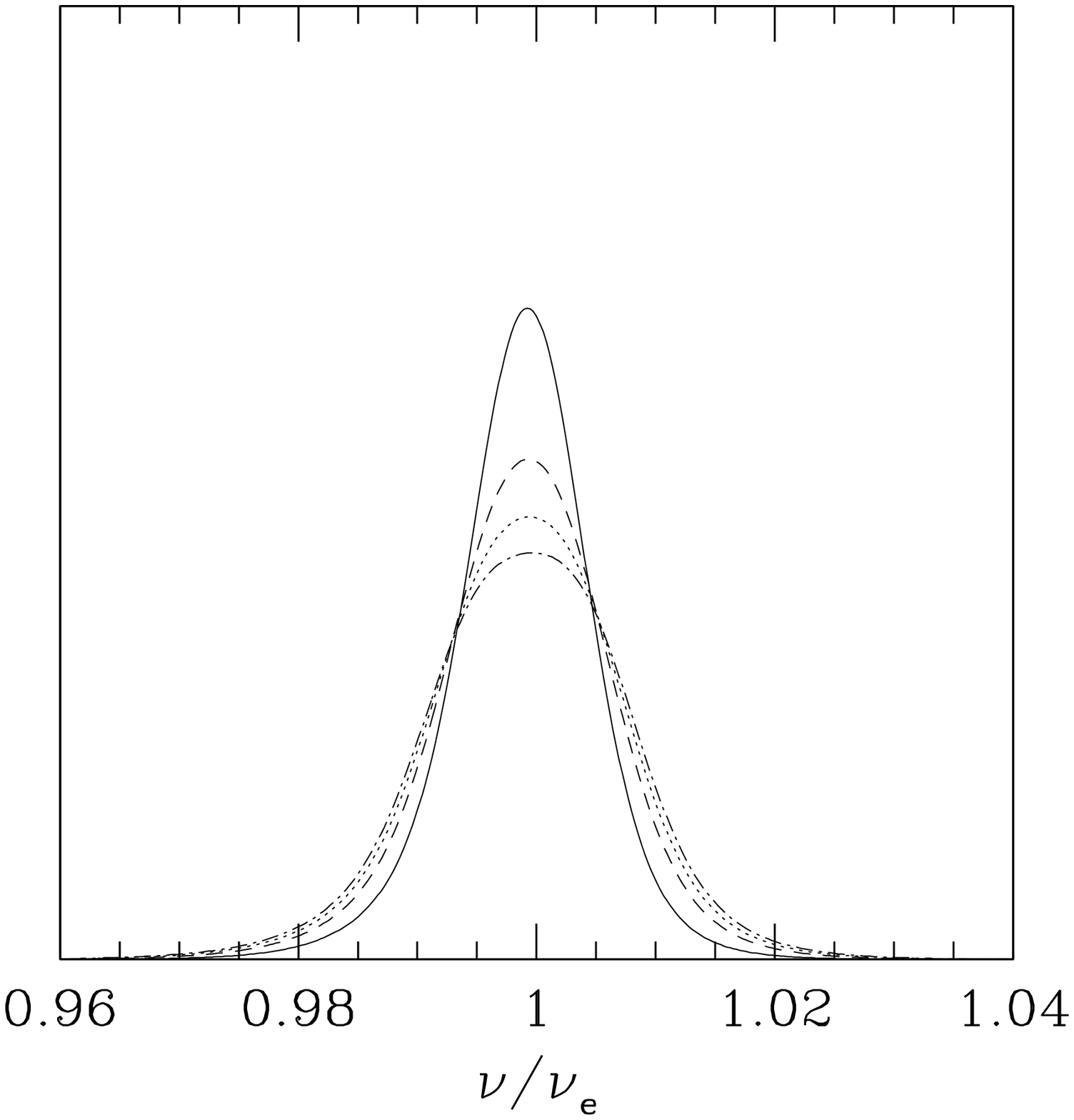,width=7.0cm,height=7.0cm}}
\caption{The profiles of thermal oxygen line for model B, with
different inclination
angles: $i=~0^\circ$(solid), 45$^\circ$(dashed), 60$^\circ$(dotted),
75$^\circ$(dash-dotted).} 
\end{figure}

\begin{figure}
\centerline{\psfig{figure=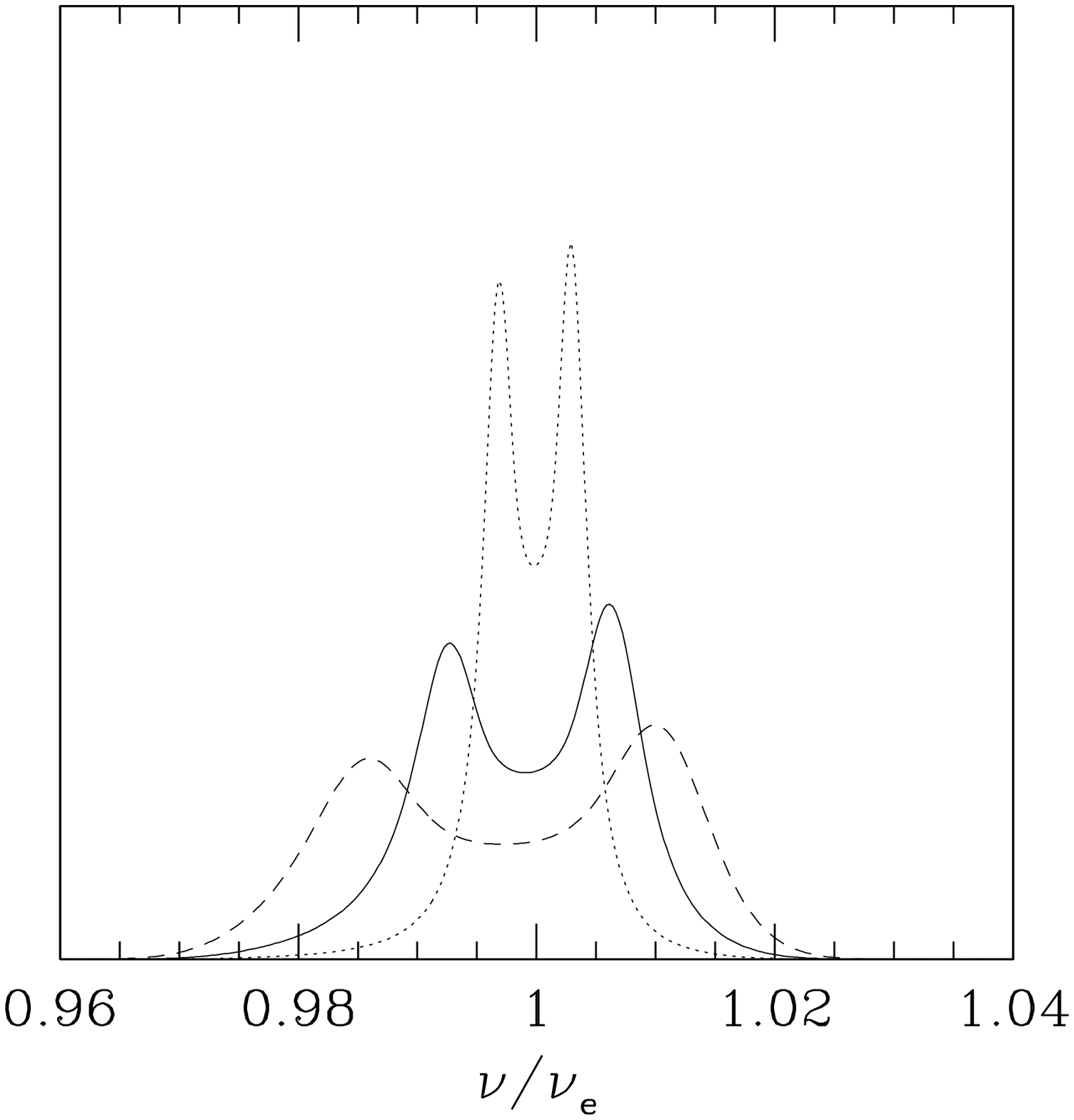,width=7.0cm,height=7.0cm}}
\caption{The line profile dependence of the outer radius for model A:  
$r_{out}=300 r_g$(dashed). $1000 r_g$(solid), $5000 r_g$(dotted). The 
inclination angle is $i=45^{\circ}$.  }
\end{figure}

\begin{figure}
\centerline{\psfig{figure=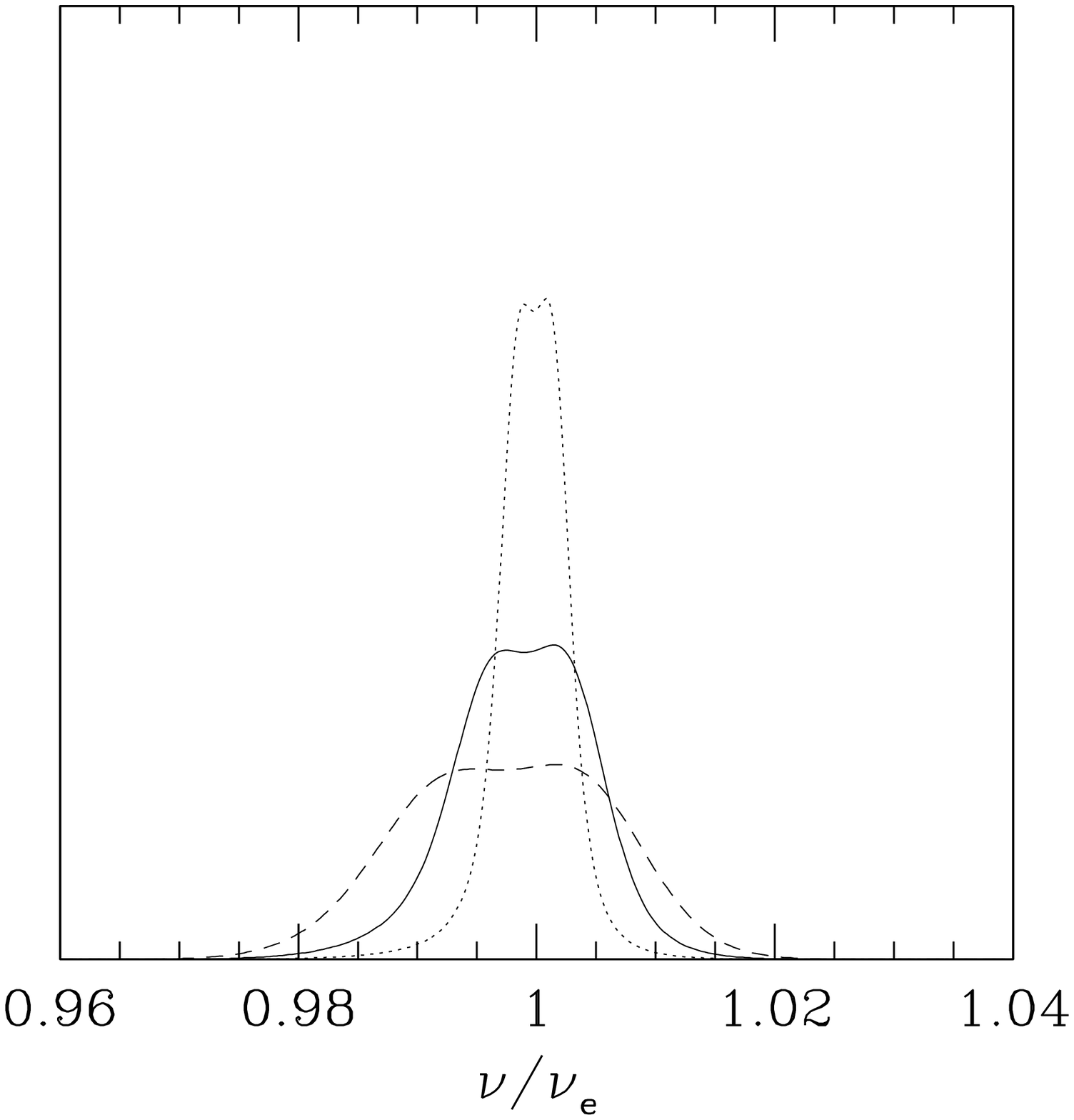,width=7.0cm,height=7.0cm}}
\caption{Same as Fig. 11, but for model B.} 
\end{figure}

\section{Results and discussion}

We depict the total bulk velocity of the flow as functions of the disk
parameters $\beta$ and $\alpha$ in Fig. 1. It is found that the velocity
is mainly determined by the value of $\beta$. The viscosity $\alpha$ affects
the velocity  very little, especially in the range of typical values of
$\alpha$ around $0.2\sim 0.3$ (Narayan et al. 1998). One may therefore
draw a conclusion that the line profile is mainly determined by the value
of magnetic pressure ratio $\beta$ that determines the ratio of specific
heats $\gamma$ here. 
We note that Quataert \& Narayan (1999) give a different relation 
between the ratio of specific heats $\gamma$ and plasma $\beta$,  
in which the particle heating is explicitly included. 
The plasma $\beta$ is a function of the radius $r$, and the structure 
of ADAFs is therefore sensitive to the outer boundary condition. 
In this work we limit our calculations to Narayan \& Yi's (1994) 
self-similar ADAF model in order to get the basic features of thermal line
profiles. The present thermal line profile calculations can be 
easily extended for different ADAF models. The observed thermal line 
profiles would be used as tests on different ADAF models, and some 
physical quantities of ADAFs can be inferred from X-ray observations 
on thermal line profile and continuum emission.

In Figs. 2--4, we compare the importance of turbulent motions 
in the formation of the line profiles for different values of viscosity 
$\alpha$. For small $\alpha$, the turbulent velocity of eddies is 
small, and the line profiles are hardly affected by turbulence, 
while the profiles are obviously broadened in 
high $\alpha$ cases (see Fig. 4).

We adopt two sets of the disk parameters: A. $\alpha=0.2$, $f=1$,
$\beta=0.5$; B. $\alpha=0.2$, $f=1$, $\beta=0.9$. 
The ratio of specific heats is $\gamma\simeq 1.444$
for $\beta=0.5$, while $\gamma\simeq 1.602$ for $\beta=0.9$.
The line profiles for models A and B are plotted in Figs. 5 and 6, 
respectively. The exponent of the emissivity law is assumed to be 
$b=1$ in these calculations. We find that
the double-peaked profiles appear for high inclination angles and
are smeared by the thermal and turbulent broadening in low 
inclination cases. The double-peaked structure finally disappears 
when the line of sight is close to the axis of the disk. The bulk velocity 
of accretion flows is large in the low-$\beta$ case, and one can then find
in Figs. 5 and 6 that the wings of line profiles for model A extend over a
larger range of frequency than that for model B. The double-peaked structure
disappears completely when the inclination angle is about 15$^{\circ}$
for model A, while an obvious plateau is present at the inclined angle around
45$^{\circ}$ for model B (see Fig. 6).

In Figs. 7 and 8, the line profiles corresponding to the different
emissivity law $b=0, ~1, ~2$ are plotted for models A and B respectively.
The line emission from the outer region of the
disk dominates in the case of $b=0$ and therefore has a narrower line profile,
while a broader line profile is present in the case of $b=2$, since the
line emission is mainly from the inner region of the disk in this case
where both the bulk and thermal velocities of the flow are high.

The line profile calculations in this work are mainly performed for
iron, which 
would be most probably observed by X-ray observatories. 
We also plot the thermal oxygen line profiles in Figs. 9 and 10 for models
A and B, respectively. It is found that the double-peaked structures 
are smeared even for high inclination cases. For model B
($\gamma=1.602$), the bulk velocity of ADAF is relatively low, the 
double-peaked structure does not appear for any inclination angle, 
and only a plateau is present for the high inclination angle (see Fig. 10).
The thermal lines for light ions have systematically broad profiles
due to the fact that the lighter ion has higher thermal motion
velocity for the same ion temperature. The expected difference between 
line profiles for different ions can be tested by future X-ray observations.

It is proposed that a transition radius $r_{tr}$ exists in the ADAF 
model. In the region of $r<r_{tr}$, the accretion flow is a
two-temperature ADAF, while in the region beyond the radius $r_{tr}$, 
the accretion occurs partially as a thin disk (Narayan, McClintock 
\& Yi 1996). The profiles of thermal line will
be obviously affected by the transition radius $r_{tr}$, 
which may be related to the accretion
rate $\dot m$ (Narayan \& Yi 1995b). 
In present line profile calculations, we have fixed the outer radius
of line emission
region as $1000 r_g$. We depict the line profile dependence of the 
outer radius in Figs. 11 and 12. It is found that the line profile 
becomes narrow and a small difference between two peaks is present 
for a large outer radius. The outer region of the ADAF has relatively 
lower bulk velocity, and the emission from this region contributes 
mainly on the central part of the profile, which leads to a narrow
line profile for a large outer radius. In the case of 
the ADAF with a small outer radius, the Doppler beaming effect 
caused by bulk motion of the flow in  the whole 
line emission region is strong, and therefore the line profile is  broad.
 
The line profiles are calculated on the assumption of the thermal distribution
for ions (but also see Mahadevan \& Quataert 1997). The thermal and 
turbulent broadening of the line is important, but the Doppler
broadening by the motion of accretion flow cannot be neglected
either. The profiles for thermal lines from ADAFs show that the blue 
peak is higher than the red one and line profiles extends more in the 
red wing, which is similar to
previous calculations on the line profiles from the standard accretion disk. 
The difference between two peaks of the line from ADAFs is smaller
than 
that of standard accretion
disk cases, and the peaks are finally smeared by the thermal and
turbulent broadening
when the inclination is low. We note that the profiles of the neighbouring
X-ray lines calculated by Narayan \& Raymond (1999), for example,
Fe\,{\sc xxv}$\lambda$1.855 and Fe\,{\sc xxvi}$\lambda$1.780, may partly
overlap in high inclination cases. The present calculations
on thermal line profiles can be used as a diagnosis on the future
X-ray observations of the thermal lines.

We have not considered the Comptonization in ADAFs in present 
calculations, which may affect the thermal line profiles to some
extent. It can be included in future line profile calculations, 
perhaps by numerical approaches (Dumont et al. 2000). 

Our line profile calculations are based on the self-similar 
height-integrated ADAF models. The vertical structures of the  
velocity, density and ion temperature have not been taken into account. 
The velocity field in ADAFs may depend on the vertical height 
(Narayan \& Yi 1995a; Igumenshchev \& Abramowicz 1999). Narayan \& Yi 
(1995a) studied the structure of ADAFs in $r\theta$ plane by 
numerically solving the self-similar equations. They compared 
their exact self-similar solutions with those obtained using 
the system of height-integrated equations and found that various 
dynamical variables such as the radial velocity, angular velocity and 
sound speed estimated from height-integrated approximate solutions 
agree very well with the corresponding spherically averaged quantities
in their exact solutions. Their exact solutions also show the
angular velocity almost not varying with the angle $\theta$ at given 
radius $r$ for small $\epsilon^{\prime}$. The present calculations can
therefore be good approximations of the reality, 
provided the emissivity is interpreted as that from the shell at given
radius $r$. We believe that the main features of line profiles have not been 
affected much by the limitation of one-dimensional velocity structure 
adopted in the calculations. The further calculations considering 
$r\theta$ distribution of variables $v_r$, $v_\phi$, $\rho$ 
and ion temperature $T_i$ are necessary for detailed modelling 
on line profiles.

\acknowledgements {We thank the referee's helpful comments. 
XC thanks Suzy Collin and Xuebing Wu for helpful discussion. 
This work is supported by NSFC(No. 19703002), the Major State Basic 
Research Development Project, and Pandeng Project. }

{}

\end{document}